\documentclass[10pt,twocolumn]{IETEJR}
\usepackage{geometry}
\usepackage{color}
\usepackage{wrapfig}
\usepackage{graphicx}
\usepackage{caption}
\usepackage{subcaption}
\usepackage{amsmath}
\usepackage{amsthm}
\usepackage{amssymb}
\usepackage{amsfonts}
\usepackage{mathtools}
\usepackage{authblk}
\usepackage{cite}
\makeatletter
\def\BState{\State\hskip-\ALG@thistlm}
\makeatother
\usepackage{listings}
\usepackage{tikz}
\usetikzlibrary{arrows,matrix,positioning}
\newtheorem{theorem}{Theorem}

\makeatletter
\renewcommand\@biblabel[1]{#1.}
\makeatother

\renewcommand{\vec}[1]{\ensuremath{\mathbf{#1}}}

\newcommand{\Figref}[1]{Figure~\ref{fig:#1}}
\newcommand{\secref}[1]{Section~\ref{sec:#1}}
\newcommand{\thmref}[1]{Theorem~\ref{thm:#1}}

\begin{document}

\title{\vspace{-8mm} \huge \normalfont Analysis of Beam Sweeping Techniques for Cell-Discovery in mm Wave Systems}
\author[1]{Rashmi P}
\author[2]{Manoj A}
\author[3]{Arun Pachai Kannu}
%\author[2]{Corresponding Author\thanks{email@2nduniversity.com}}
\affil[1,2,3]{Electrical Engineering Department,IIT Madras, Chennai - 600036, India.}
{
    \makeatletter
    \renewcommand\AB@affilsepx{, \protect\Affilfont}
    \makeatother
    \affil[1]{ee15d054@ee.iitm.ac.in}
    \affil[2]{ee14d210@ee.iitm.ac.in}
    \affil[3]{arunpachai@ee.iitm.ac.in}
}
\date{}
\twocolumn[
\begin{@twocolumnfalse}
    \maketitle
    \vspace{-5mm}
    \begin{abstract}
    Cell discovery is the procedure in which a user equipment (UE) finds a suitable base station (BS) to establish a communication link. When beamforming with antenna arrays is done at both transmitter and receiver, cell discovery in mm wave systems also involves finding the correct angle of arrival  - angle of departure  alignment between the UE and the detected BS. First, we consider the single BS scenario and present the mathematical model for the cell discovery problem. We present the details of the beam sweeping based techniques and explain the beam combining scheme to reduce the training overhead. We analytically characterize the performance of energy detector at the UE for both beam sweeping and beam combining methods under some channel assumptions. Further, we consider the multiple BS case where BSs transmit synchronization signals along with directional beamforming. We analyze the performance of the energy detector in this case as well. We finally present simulation studies with channels generated using generic mm wave channel emulators and draw inferences on the role of various parameters on the cell discovery performance.
    \end{abstract}
\end{@twocolumnfalse}
\vspace{5mm}
]

\section{INTRODUCTION}
\vspace{-3mm}
\par Millimeter Wave (mm wave) communication is a
potential candidate for the 5G cellular technology to meet the continuous increase in the volume of mobile data, the demand for high data rate services and to achieve large network coverage 
\cite{Xiao_2017_JSAC,Heath_2014_TWC,Zhang_2014_GLOBE,Mukesh_IETE_2019}. As revealed by various studies carried out on different path loss models \cite{Akdeniz_2014_JSAC, Rappaport_2017_ICC}, mm wave systems suffer from poor propagation characteristics. 
At the same time, the small value of the carrier wavelength enables deployment of a large number of antennas packed into small sized arrays at the communicating nodes, which in turn helps in achieving high spatial gains via beamforming \cite{Daasari_IETE_2021}. Cell discovery (CD) is a procedure by which user equipment (UE) entering a network finds a suitable (nearby) base station (BS) and its identity \cite{APK_2015_TCOM}, in order to establish a link-layer connection. Because of the large path loss, unlike in the sub-6GHz systems, the initial cell discovery cannot be done using omni-directional transmission of synchronization signals in mm wave systems, as the propagation range will be highly compromised. Hence, mm wave systems rely on directional signaling for CD, making the cell discovery process very challenging as the large angular space is scanned.
%
% Recent works on antenna array designs, developments in channel estimation methods \cite{Arun_2018_TCOM, Lee_2016_TCOM, Ghauch_2016_JSTSP} and studies on different path loss models \cite{Akdeniz_2014_JSAC,Rappaport_2017_ICC} have further increased the feasibility of deploying mmwave systems for short range indoor and outdoor wireless communication. However, directional signal transmission and reception is preferred over mm wave channels to compensate for the high path loss. As a result, unlike in the sub-6GHz systems, the initial cell discovery cannot be done using omni-directional transmission of synchronization signals in mm wave systems.
%
%User discovery is a dual problem of CD, where a BS tries to identify the presence of a UE intending to make a connection. 
%Several studies have been done in the literature to address cell or user discovery problem in mm wave systems. 
Beam sweeping is a conventional exhaustive CD technique \cite{Barati_2016_TWC,Hur_TCOM_2013}, where both transmitter and receiver sweep their beams along the entire angular range (say, 360 degrees) and make a detection based on the angle of arrival (AoA) - angle of departure (AoD) beam-pair corresponding to the largest received signal strength. Beam sweeping scheme has the simplest form of directional beamforming codebook design and guarantees very good cell discovery performance. Directional beamforming codebook-based beamforming was discussed in \cite{Raghavan_2016_JSTSP} in which the authors validate the importance of directional beamforming for practical mm wave networks. Several variations of beam sweeping schemes have been proposed in the literature. In \cite{Barati_2015_TWC}, a detection algorithm with omni-directional transmission and beam sweep based reception is shown to have a better trade-off over the directional beam sweep in terms of received SNR versus detection delay. Nevertheless, the omni-directional stage compromises the transmission range of the set up. Beam sweeping with frequency multiplexed beams is discussed in \cite{Desai_ASIL_2014} with an extension on how to combine the beams in the frequency domain to decrease the number of search beams. Another sub-class of beam sweep is the iterative CD technique \cite{Giordani_2016_CISS,Habib_2017_IWCMCC}. Here the search for accessible beams starts with wider sector shaped beams, and the beams are then narrowed and refined based on the continuous feedback of signal strength. However, the feedback overhead from UE or BS during the CD process scales with the number of UEs. In addition, BS has to adapt its transmission beams based on the UE feedback, which implies that the training process of each UE needs to be carried out separately. In \cite{Habib_2017_IWCMCC} a hybrid cell search method is proposed, which combines the important aspects of the exhaustive search and the iterative search techniques. CD schemes with context information are proposed in  \cite{Filippini_2018_TMC, Devoti_2016_IEEEAccess, Rezagah_2015_CSCN,Alex_2017_CAMSAP}. However, the performance of these schemes completely relies on the accuracy of the estimate on user position.

\par When multiple BSs are present in a wireless network, synchronization sequences are transmitted by BSs to enable accessible base station identification at the UE, along with the best beam identification. In 3GPP-NR, sequences used by the BSs are termed as primary/secondary synchronization signals (PSS/SSS) \cite{Barati_2015_TWC,Giordani_ICST_2019}. Each BS will have a dedicated PSS/SSS, and these are typically constructed using Zadoff-Chu (ZC) or m-sequences which have good correlation properties. However, theoretical guarantees on the CD performance of the beam sweep methods were not well established in the literature to the extent of our knowledge.

\par Our main contributions in this paper are summarised below. 
\begin{itemize}
    \item We elaborately discuss the physical layer implementation of the CBS scheme by completely specifying its training phase, with the explicit description of beamforming vectors to be used at the transmitter and receiver. We also discuss how the beamforming vectors can be modified to implement beam widening or beam combining variations (Beam Combining (BC) method) with the explicit characterization of the training phase.
    \item We derive analytical expressions/bounds for the detection probability and false alarm probability for the CBS and BC methods, under certain assumptions on the mm wave channel models.  
    \item We analyze the time delay associated with CD and also the average number of repeated attempts required for CD mathematically and derive analytical expressions for the same under special channel conditions.
    \item We also discuss the beam sweeping scheme for multiple BS case using sequence based transmission and characterize the detection probability and false alarm probability when BSs use orthogonal sequences during training. 
    \item We evaluate and compare the performance of CD considering generic wireless mm wave channel model generated using NYUSIM simulator (Version 1.6) \cite{Rappaport_2017_ICC,Rappaport_2017_VTC}. The simulator being developed from extensive real world experiments, gives an insight into how well the CBS and BC schemes perform in outdoor environment. 
\end{itemize}

{\emph{Notations:}} Mathcal font $\mathcal{P}$ denotes a set and its cardinality is $|\mathcal{P}|$. $p$ and $P$ denote scalars, $\vec{p}$ - vector and $\vec{P}$ - matrix. $[\vec{P}]_{p,\bar{p}}$ refers to the $(p,\bar{p})^{th}$ element of matrix $\vec{P}$ and $\vec{p}_p$ the $p^{th}$ element of vector $\vec{p}$. Also, $\vec{p}_{\mathcal{P}}$ is a sub-vector of $\vec{p}$ containing elements of $\vec{p}$ indexed by $\mathcal{P}$. $p^{th}$ row and $p^{th}$ column of $\vec{P}$ is denoted as $[\vec{P}]_{p,:}$ and $[\vec{P}]_{:,p}$ respectively. $\vec{I}_p$ and $\vec{F}_p$ denote identity and unitary DFT matrices of size $p \times p$ each. $\vec{0}$ indicates zero matrix, and $\vec{1}_p$ indicate an all-one vector of length $p$. Lastly, $||.||_2$ is the $l_2$-norm and $\mathbb{E}[.]$ is the expectation operator. $(\cdot)^*$ and $(\cdot)^T$ denote Hermitian and transpose respectively. 

\vspace{-2mm}

\section{System Model and Problem Statement}
\label{sec:sys_model}
\label{sec:prob_form}
In this section, we present the details of the cell discovery problem in mm wave systems for the single base station scenario. The case of multiple base stations is addressed in 
\secref{sequence transmission}
\vspace{-5mm}
\subsection{Channel Model} \label{sec:chan}
\par Consider a mm wave communication system transmitter, which we refer as a base station (BS), equipped with a uniform linear array of antennas (ULA) of size $N_t \times 1 $. Assume that the receiver, which we refer as the user equipment (UE), has a ULA of size $N_r \times 1$. The MIMO channel from the BS to the UE, denoted by $\vec{H}$ (of size $N_r \times N_t$), is modeled as \cite{Heath_2014_TWC,Lee_2016_TCOM,Arun_2018_TCOM},
\begin{align}
\vec{H} = \sum\limits_{k = 1}^{K} \alpha_{k} \vec{a}_{r_k} \vec{a}_{t_k}^*,
\label{channel_model}
\end{align}
where $K$ is the total number of multi-path components, $\alpha_{k}$ is the complex channel gain of the $k^{th}$ multi-path and $\vec{a}_{t_k}$ and $\vec{a}_{r_k}$ are the corresponding transmit and receive antenna array response vectors.
The array response vectors depend on the AoA and AoD of the corresponding multi-path. Specifically, 
$\vec{a}_{r_k} = \frac{1}{\sqrt{N_r}}[1 \, e^{-j\omega_{r_k} } \, ... \, e^{-j(N_r-1)\omega_{r_k}}]^T$ and $\vec{a}_{t_k}  = \frac{1}{\sqrt{N_t}}[1 \, e^{-j\omega_{t_k} } \, ... \, e^{-j(N_t-1)\omega_{t_k} }]^T$,
%\begin{align*}
%\vec{a}_{r_k}  = \frac{1}{\sqrt{N_r}}[1 \, e^{-j\omega_{r_k} } \, ... \, e^{-j(N_r-1)\omega_{r_k} }]^T; \\ \vec{a}_{t_k}  = \frac{1}{\sqrt{N_t}}[1 \, e^{-j\omega_{t_k} } \, ... \, e^{-j(N_t-1)\omega_{t_k} }]^T.
%\end{align*}
where $\omega_{r_k}  = 2\pi \frac{d}{\lambda} \sin(\theta_{r_k} )$ and $\omega_{t_k}  = 2\pi \frac{d}{\lambda} \sin(\theta_{t_k} )$, $d$ is the inter-element spacing between adjacent antennas in the ULA (at both the BSs and the UE), $\lambda$ is the operating carrier wavelength, and $\theta_{r_k} $ and $\theta_{t_k} $ are the AoA and AoD respectively, associated with the $k^{th}$ multi-path component of the channel $\vec{H}$. The channel gains, $\alpha_{k} $'s, are distributed as independent $\mathcal{CN}(0,\sigma_{k}^2)$. % with $\sigma_{k_i}^2$ being nor $ = \sqrt{\frac{N_tN_r}{K_i}}\tilde{\sigma}_{k_i}^2 $. This is to ensure $\mathbb{E}[||\vec{H}_i||_F^2] = N_tN_r,\, \forall i$ \cite{Heath_2014_TWC}. 

\par Each AoA-AoD pair $(\theta_{r_k} ,\theta_{t_k} )$ leads to a spatial frequency pair $(\omega_{r_k} ,\omega_{t_k} )$ so that the 2D Fourier transform of $\vec{H}$ will have significant energies in the DFT bins closer to $(\omega_{r_k} ,\omega_{t_k} )$. Since the number of multi-paths is small compared to the array sizes, the mm wave channels are approximately sparse in Fourier basis \cite{Mo_ASIL_2014,Lee_2016_TCOM,Arun_2018_TCOM}. Specifically, 2D Fourier transform of $\vec{H}$, computed using unitary DFT matrices as,
\begin{align}
\vec{G} = \vec{F}_{N_r}^* \vec{H} \vec{F}_{N_t},
\label{sparse_ch_rep}
\end{align}
is an approximately sparse matrix. The locations of significant valued entries in $\vec{G}$ provide indication of 
the AoA-AoD pairs corresponding to the strong paths between UE and BS. %\Figref{sparse_channel_magnitude} shows the 2-D magnitude plot of  $\vec{G}$ matrix with $N_t=N_r=32 \text{ and } K=4$. The green-yellow coloured squares indicates the spatial frequency bins corresponding to the channel paths \cite{Arun_2018_TCOM}.
In the special case of the spatial frequencies falling exactly on the DFT bins (which we refer as \emph{ideal channel conditions}),   
\begin{eqnarray}
\omega_{r_k}  &\in& \Big\{\frac{2\pi l}{N_r}, l = 0,\cdots,N_r-1 \Big \}, \forall k, \label{ideal1}\\
\omega_{t_k}  &\in& \Big\{\frac{2\pi m}{N_t}, m = 0,\cdots,N_t-1 \Big \}, \forall k, \label{ideal2}
\end{eqnarray}
the matrix $\vec{G}$ is an \emph{exactly} sparse matrix with $K$ independent Gaussian non-zero entries (each with a variance of $\sigma_k^2$) in the DFT bins given by the spatial 
frequency pairs $(\omega_{r_k} ,\omega_{t_k} )$. However, in a practical set-up, the AoA-AoD values will be random and the $\vec{G}$ matrices will be approximately sparse.

\vspace{-2mm}
\subsection{Beamforming Training}
\par Suppose the BS transmits data symbol $x$ using a beamforming vector $\vec{w}_{t}$, then the received signal at the UE $y$ is given by,
\begin{align}
y &= \sqrt{\rho}\vec{w}_r^* \vec{H} \vec{w}_{t}x + \underbrace{\vec{w}_r^* \vec{n}}_{n},
\label{sys_model}
\end{align}
where $\rho$ is the transmit power, $\vec{w}_r$ is the receive beamforming vector and $\vec{n}$ is the additive $N_r$-length noise vector. We assume, $\vec{n} \sim \mathcal{CN}(\vec{0}, \sigma_n^2 \vec{I}_{N_r})$ and hence $n \sim \mathcal{CN}(0,\sigma_n^2 ||\vec{w}_r||_2^2)$. %We strictly adhere to the system model stated in equation \eqref{sys_model} and 
For convenience and consistency, we always set the receive beamforming vector to be of unit norm, which yields $n \sim \mathcal{CN}(0,\sigma_n^2)$.
%If $||\vec{w}_r||_2^2 = 1$, then the noise for each scalar measurement can just be modeled as a single Gaussian random variable (instead of a vector). Hence, we always ensure that the receive beamforming vector is always of unit norm and we strictly adhere to the system model stated in equation \eqref{sys_model}.

%\section{Problem Formulation}
%\label{sec:prob_form}
\par In this paper, we consider the cell discovery in mm wave systems where a UE endeavors to detect the presence of the BS to establish a communication link. %Identifying a BS implies retrieving the BS's identity and the associated AoD-AoA alignment. 
For this purpose, we consider the training phase where the BS and UE use known/fixed set of beamforming vectors and pilot symbols. 
UE has a set of $P$ beamforming vectors $\{\vec{w}_r^{(1)},\cdots,\vec{w}_r^{(P)}\}$ and the BS has a set of $Q$ beamforming vectors 
$\{\vec{w}_{t}^{(1)},\cdots,\vec{w}_{t}^{(Q)}\}$. The received observation corresponding to a given transmit/receive beamforming vectors pair is,
\begin{eqnarray}
y_{p,q} &=& \sqrt{\rho} \vec{w}_r^{(p)*} \vec{H} \vec{w}_{t}^{(q)}x_{p,q} + n_{p,q},
\label{pq_meas_eq} \\ 
&=& \sqrt{\rho} \vec{w}_r^{(p)*}  \vec{F}_{N_r} \vec{G} \vec{F}^*_{N_t} \vec{w}_{t}^{(q)}x_{p,q} + n_{p,q}, \label{pq_meas_g} 
\end{eqnarray}
where $x_{p,q}$ are the known pilot symbols and $\{n_{p,q}\}$ are i.i.d. complex Gaussian noise samples with variance  $\sigma_n^2$. We assume that all the pilot symbols have unit magnitude, $|x_{p,q}|= 1,\, \forall p,q$. 
%Also, without loss of generality, we assume that all the BSs use same transmit power during the training phase, so that, for each $q = 1, ..., Q$, $||\vec{w}_{t_i}^{(q)}||_2^2$ must be equal $\forall i = 1,\cdots,N_{\text{bs}}$. 
If all the combinations of transmit/receive beamforming pairs are used, then the training phase duration will be $M=PQ$. For subsequent use, we define the received SNR of the training phase as,
 \begin{equation}
\text{SNR} = \frac{\sum_{p,q} \big|\sqrt{\rho}~\vec{w}_r^{(p)*} \vec{H} \vec{w}_{t}^{(q)}x_{p,q} \big|^2}{M\sigma_n^2}. \label{snr1}
\end{equation}
 
\subsection{Cell Discovery Problem}
In the hypothesis testing framework for the cell discovery problem, if the UE receives signal from the BS, then the channel matrix $\vec{H}$ in \eqref{pq_meas_eq} 
is of the form \eqref{channel_model}. On the other-hand, if the UE does not receive the signal from the BS (when the BS is far away or got blocked by obstacles), the
channel matrix is zero $\vec{H} = \vec{0}$. The cell discovery problem involves finding the presence (or absence) of BS before initiating the communication link. If the 
presence of BS is detected, mm wave cell discovery involves finding additionally the AoA-AoD pairs 
of strong paths with respect to the identified BS.   
Specifically, using the $M$ training phase observations in \eqref{pq_meas_eq}, the UE tries to identify the presence (or absence) of the BS and also
find the significant entries in the (Fourier domain) channel matrix $\vec{G}$, which will give the required AoA-AoD pairs for the detected BS.
\vspace{-2mm}
\section{Beam Sweeping Based Techniques}
\label{sec:Conventional_tech}

\par In this section, we present the conventional beam sweeping (CBS) technique \cite{Barati_2016_TWC,Hur_TCOM_2013} which is being adopted in many wireless standards \cite{Giordani_ICST_2019} for the cell discovery problem.  In this method, the transmitter and receiver exhaustively search the optimal AoA-AoD pair, by sweeping their beams in the entire $[0,2\pi]$ space and using all the transmit/receive beam combinations. % with beams constructed using columns of unitary DFT matrices. We also mathematically characterize the detection performance of the same. However, the resultant delay in establishing a link using the beam sweep method is high. Thus, we discuss different variations of this method to reduce the overhead involved. 
We employ an energy detector to detect the presence of BS and find the corresponding AoA-AoD pairs. Under the idealized on-grid channel conditions, we analytically characterize the detection and false alarm probabilities of the CBS training scheme with the energy detector. We also present and analyze the beam combining (BC) or beam widening technique \cite{Raghavan_2016_JSTSP}, where we widen (by combining multiple beams) the beams used during the training phase, so that the overall duration to sweep the entire angular space is reduced.

\subsection{Conventional Beam Sweep:}
\label{sec:BS}
\par In this scheme, the beamforming vectors $\Big\{\vec{w}_{t}^{(q)}\Big\}_{q=1}^Q$ and $\Big\{\vec{w}_r^{(p)} \Big\}_{p=1}^P$ are chosen as 
columns of the unitary DFT matrices $\vec{F}_{N_t}$ and $\vec{F}_{N_r}$ respectively. Setting $P = N_r$ and $Q = N_t$, we 
have, %$\vec{w}_{t}^{(q)} = [\vec{F}_{N_t}]_{:,q}, \,\, q \in \{1, ..., N_t \},$ and $\vec{w}_{r}^{(p)} = [\vec{F}_{N_r}]_{:,p}, \,\, p \in \{1, ..., N_r \}$.
\begin{align}
& \vec{w}_{t}^{(q)} = [\vec{F}_{N_t}]_{:,q}, \,\, q \in \{1, ..., N_t \}, 
\label{simplified_weight_vectors_1} \\&
\vec{w}_{r}^{(p)} = [\vec{F}_{N_r}]_{:,p}, \,\, p \in \{1, ..., N_r \}.
\label{simplified_weight_vectors_2}
\end{align}
\par Note that when a beamforming vector is set as a column of a DFT matrix, the signal radiated by the antenna array is directed at an angle corresponding to the (spatial) frequency given by that column. In the CBS, we perform exhaustive search by considering all the transmit/receive beam combinations, leading to a total of $M=N_tN_r$ measurements, which are given by,
%On substituting equation \eqref{sparse_ch_rep} in equation \eqref{pq_meas_eq}, we get,
\begin{align}
y_{p,q} & = \sqrt{\rho} [\vec{F}_{N_r}]_{:,p}^* \underbrace{\vec{F}_{N_r} \vec{G} \vec{F}_{N_t}^*}_{\vec{H}} [\vec{F}_{N_t}]_{:,q} x_{p,q} + n_{p,q} \nonumber \\ 
& =  \sqrt{\rho} x_{p,q} [\vec{G}]_{p,q} + n_{p,q},
\label{simplified_obs_model_BS}
\end{align}
with $p\in\{1,..., N_r\},~~q\in\{1,...,N_t\}$. As noted earlier, the Fourier domain channel matrix $\vec{G}$ is approximately sparse. We define the support set $\mathcal{S}$ of $\vec{G}$ as,
\begin{align}
\mathcal{S}= \Big\{(a,b) \Big| |[\vec{G}]_{a,b}|^2 > \delta, \, a = 1, ..., N_r;b = 1, ..., N_t \Big\},
\label{support_set}
\end{align}
where $\delta$ is an appropriately chosen limit value that declares whether or not an entry in matrix $\vec{G}$ has considerably large magnitude. For $(p,q) \in \mathcal{S}$, the received symbols $y_{p,q}$ in \eqref{simplified_obs_model_BS}, have significant signal component due to sufficiently large channel gains and we term them as \textit{active symbols}. For $(p,q) \notin \mathcal{S}$, the channel gains are close to zero and the corresponding observations are noise dominated. 
%Suppose we define an index set $\mathcal{S}_{\text{actv}}$ as,
%\begin{align*}
%\mathcal{S}_{\text{actv}} &= \Big\{ (p,q) \Big| y_{p,q} \, \text{is generated by} \, [\vec{G}_i]_{\bar{p},\bar{q}_i}, \, \\ &
%\hspace*{40mm}  \text{for some} \, (\bar{p},\bar{q}_i) \in \mathcal{S} \Big\},
%\end{align*}

Now, detecting the presence of at least one accessible BS is equivalent to verifying whether there exists at least one active symbol among $\Big\{ y_{p,q}, p \in \{1,..., N_r\},~~q\in\{1,...,N_t\} \Big\}$. %But during the cell discovery phase, the UE will not have any information about the channels $\vec{H}_i$ (and so about the matrices $\vec{G}_i$ and the index set $\mathcal{S}$). Hence, we propose the following algorithm: the UE can declare that an accessible BS is present in the network, if there exists at least one received symbol $y_{p,q}$ such that,
In other words, the BS is detected with AoA-AoD pair given by the angular frequencies $\hat{\omega}_t = \frac{2\pi p}{N_t}, \hat{\omega}_r = \frac{2\pi q}{N_r}$, if 
\begin{align}
 |y_{p,q}|^2 > \tau,
 \label{thr_condition}
\end{align} 
where $\tau$ is an appropriately chosen threshold. If threshold condition \eqref{thr_condition} is not satisfied for any $(p,q)$ pair, the detector declares that BS is absent. %The basic idea is, suppose $y_{p_0,q_0}$ for some $p_0 \in \{1,...,N_r\}$ and $q_0 \in \{1,...,N_t\}$ satisfies equation \eqref{thr_condition}, then it is expected that the training vectors $\vec{w}_r^{(p_0)}$ and $\vec{w}_{t_i}^{(q_0)}$ might have generated the  transmit-receive beams that are actually aligned along an existing channel path. Since only the active symbols are the ones that are generated by the large magnitude elements of $\vec{G}_i$'s (i.e., the channel gains), it is crucial to ensure that there is very less chance for any noise dominated received symbol to satisfy \eqref{thr_condition}. If \textit{at least one received symbol not indexed by the set $\mathcal{S}$ satisfies equation \eqref{thr_condition}}, then we term such an event as false alarm. The probability of false alarm ($\mathbb{P}_F$) is then defined as,
%The indices $(p,q)$ corresponding to the active symbols indicate the AoA-AoD alignment between the BS and UE. 
For the above detection rule \eqref{thr_condition}, the probability of false alarm $\mathbb{P}_F$ and the probability of successful detection $\mathbb{P}_D$ are given by
\begin{eqnarray}
\mathbb{P}_F &=& \mathbb{P}(\bigcup\limits_{(p,q) \notin \mathcal{S}} |y_{p,q}|^2 > \tau ), % \nonumber \\ &=& \mathbb{P}(\max\limits_{(p,q) \notin \mathcal{S}} |y_{p,q}|^2 > \tau ), 
\label{Pf_defn_BS} \\
\mathbb{P}_D &=&  \mathbb{P}\Big(\bigcup\limits_{(p,q) \in \mathcal{S}} |y_{p,q}|^2 > \tau \Big). \label{Pd_defn_BS}
\end{eqnarray}
%The threshold $\tau$ should be chosen such that $\mathbb{P}_F$ is low. 
%The symbols $y_{p,q}, \, \forall (p,q) \in \mathcal{S}$ are, in general, correlated because the matrices $\Big\{ \vec{G}_i \Big\}_{i =1}^{N_{\text{bs}}}$ are only approximately sparse. Hence, in order to analyze and appreciate the $\mathbb{P}_D$ performance mathematically, we consider the \emph{idealized channel assumption}, described in Appendix A.

%As a consequence, the expression for the received symbols defined in equation \eqref{simplified_obs_model_BS} can be simplified as,
%\begin{equation}
%y_{p,q} = \begin{cases}
%s_{p,q} + n_{p,q}, & \, \text{if} \, (p,q) \in \mathcal{S}_{\text{actv}},\\
%n_{p,q}, & \, \text{if} \, (p,q) \notin \mathcal{S}_{\text{actv}}
%\end{cases},
%\label{refraces_ypq_BS}
%\end{equation}
%where $s_{p,q}$ is the signal part of the active symbol $y_{p,q}, \, (p,q) \in \mathcal{S}_{\text{actv}}$, and is a function of certain number of non-zeros entries from matrices $\Big\{\vec{G}_i \Big\}_{i=1}^{N_{\text{bs}}}$ (at least one of them, but not necessarily all of them). This implies, $|\mathcal{S}_{\text{actv}}| \leq |\mathcal{S}|$. It is important to note that the active symbols are independent, but due to the uncertainity about the structure of $s_{p,q}, \, \forall (p,q) \in \mathcal{S}_{\text{actv}}$, it is impossible to determine the exact statistics of the active symbols. Because of which, we cannot derive an exact expression for $\mathbb{P}_D$, but we could derive a lower bound for it as stated in the following theorem.

\begin{theorem} \label{thm:cbs}
For the conventional beam sweep technique % with  \eqref{simplified_weight_vectors_1} and \eqref{simplified_weight_vectors_2} and $x_{(p,q)}^{(i)} = \sqrt{\rho} ,\, \forall i $, then 
with the energy detector \eqref{thr_condition}, under the \textit{idealized channel conditions} given in \eqref{ideal1} and \eqref{ideal2}, 
the probability of false alarm is given by
\begin{equation}
\mathbb{P}_F = 1-\Big(1-e^{\frac{-\tau}{\sigma_n^2}}\Big)^{N_t N_r - |\mathcal{S}|}. \label{thres_expression}
\end{equation}
For the same detector, the probability of successful detection $\mathbb{P}_D$ is given by 
\begin{align}
\mathbb{P}_D = 1- \prod \limits_{k=1}^{K} \Big[1-\exp \Big(-\frac{\tau}{\sigma_n^2 + \rho \sigma_{k}^2} \Big)\Big].
\label{Pd_BS}
\end{align}
%where $\tau$ is a threshold chosen appropriately to meet a required false alarm probability $\mathbb{P}_F$ as,
%\begin{align}
%\tau = \sigma_n^2\log\Big(\frac{1}{1-(1-\mathbb{P}_F)^{\frac{1}{N_tN_r-|\mathcal{S}|}}}\Big),
%\label{thres_expression}
%\end{align}
%\item \textbf{Overlap case}: If $\bigcap\limits_{i=1}^{N_{\text{bs}}} \mathcal{S}_{i} \neq \emptyset$, then $\mathbb{P}_D$ can be lower bounded as,
%\begin{align}
%\mathbb{P}_D \geq \exp\Big(-\frac{\tau}{\sigma_n^2 + \rho \Big[\max\limits_{1 \leq i \leq N_{\text{bs}}} \max\limits_{1 \leq k \leq K_i} \Big\{ \sigma_{k_i}^2  \Big\} \Big]}\Big).
%\label{Pd_RBS}
%\end{align}
%where the threshold $\tau$ is same as equation \eqref{thres_expression}.
%\end{itemize}
\end{theorem}
\begin{proof}
Recall that, under the \textit{idealized channel assumption}, 2D Fourier domain channel matrix $\vec{G}$ will have exactly $K$ non-zero bins. 
The variance of the non-zero entry in a given DFT bin will be equal to the variance $\sigma_{k}^2$ of the path whose spatial frequency pair matches with the given bin frequency pair. 
%\par Since $\bigcap\limits_{i=1}^{N_{\text{bs}}} \mathcal{S}_{i} = \emptyset$, 
The observations $ \{y_{p,q} \}$ from equation \eqref{simplified_obs_model_BS} are given by, 
\[
y_{p,q} = \begin{cases}
\sqrt{\rho} x_{p,q} [\vec{G}]_{p,q} + n_{p,q}, & \, \text{if} \, (p,q) \in \mathcal{S}, \\
n_{p,q}, & \text{if} \, (p,q) \notin \mathcal{S}
\end{cases}.
\]
Note that noise samples $\{n_{p,q}\}$ are i.i.d. Gaussian with variance $\sigma_n^2$ and $\mathbb{P}(|n_{p,q}|^2 \leq \tau) = \Big(1-e^{-\frac{\tau}{\sigma_n^2}}\Big)$. We have, 
\begin{align*}
\mathbb{P}_F &= \mathbb{P} \Big(\bigcup\limits_{(p,q) \notin \mathcal{S}} |y_{p,q}|^2 > \tau \Big)\\ &= 1- \mathbb{P} \Big(\bigcap\limits_{(p,q) \notin \mathcal{S}} |n_{p,q}|^2 \leq \tau \Big) \\ &= 1-\Big(1-e^{-\frac{\tau}{\sigma_n^2}}\Big)^{N_tN_r-|\mathcal{S}|}.
\end{align*}
%Since $n_{p,q}$ are all i.i.d. $\mathcal{CN}(0,\sigma_n^2)$, we get,
%\begin{align*}
%\mathbb{P}_F & = 1- \prod\limits_{(p,q) \notin \mathcal{S}} \mathbb{P}(|n_{p,q}|^2 \leq \tau) \\& = 1-\Big(1-e^{-\frac{\tau}{\sigma_n^2}}\Big)^{N_tN_r-|\mathcal{S}|}.
%\end{align*}
%On rearranging the terms in the above equation, the threshold can be expressed as a function of $\mathbb{P}_F$ as in equation \eqref{thres_expression}.
For any $(p,q) \in \mathcal{S}$, we have $y_{p,q} \sim \mathcal{CN}(0,\rho \sigma_{k}^2 + \sigma_n^2)$ for some $k$. In addition,
these observations are independent.  As the result, the detection probability $\mathbb{P}_D$ is given by,
\begin{align*}
\mathbb{P}_D &=\mathbb{P}\Big(\bigcup\limits_{(p,q) \in \mathcal{S} } |y_{p,q}|^2 >\tau \Big) = 1- \mathbb{P} \Big(\bigcap\limits_{(p,q) \in \mathcal{S}} |y_{p,q}|^2 \leq \tau\Big)\\& {=} 1-\prod \limits_{(p,q) \in \mathcal{S}}\mathbb{P}(|y_{p,q}|^2 \leq \tau) \\&= 1- \prod\limits_{k \in \{1,\cdots,K\}} \Big(1-e^{-\frac{\tau}{\rho \sigma_{k}^{2} + \sigma_n^2}} \Big).
\end{align*}
\end{proof}
We note that, asymptotically as $\sigma_n^2 \longrightarrow 0$, $\mathbb{P}_D \longrightarrow 1$ and $\mathbb{P}_F \longrightarrow 0$. We point out that the assumption of \emph{ideal channel conditions} is needed only for deriving the analytical expressions for $\mathbb{P}_D$.  However, the training schemes and the detectors are also applicable for practical (off-grid) mm wave channels, which is substantiated by the simulation studies in \secref{Num_result}.

\subsection{Beam Combining method:}
\label{sec:weighted_comb}

\par We present the details of beam combining (BC) or beam widening technique, where we widen (by combining multiple beams) the beams used during the training phase. This helps to reduce the overall duration to sweep the entire angular space. A direct approach to widen the beams for scanning the space is to employ training beamforming vectors that are constructed as (weighted) linear combination of the columns of $\vec{F}_{N_t}$ and $\vec{F}_{N_r}$ matrices at the BS and UE respectively. Suppose, we combine $\beta_t$ beams at 
the transmitter and $\beta_r$ beams at the receiver, we need $P=\frac{N_r}{\beta_r}$ (widened) beams at the receiver and $Q=\frac{N_t}{\beta_t}$ (widened) beams at the transmitter. By using all the beam pair combinations, the total number of measurements needed is 
$M= \frac{N_t N_r}{\beta_t \beta_r}$. Compared to the CBS method, BC has reduced the training overhead by a factor of $\beta_t \beta_r$.

Specifically, we design the training phase beamforming vectors as,
\begin{eqnarray}
\vec{w}_{r}^{(p)} &=& \frac{\sum\limits_{l_r=1}^{\beta_r} [\vec{F}_{N_r}]_{:,(p-1)\beta_r + l_r}}{\sqrt{\beta_r}} ,p=1,\cdots,\frac{N_r}{\beta_r}, \label{bc_r} \\ 
\vec{w}_{t}^{(q)} &=& \frac{\sum\limits_{l_t=1}^{\beta_t} [\vec{F}_{N_t}]_{:,(q-1)\beta_t + l_t}}{\sqrt{\beta_t}}, q=1,\cdots,\frac{N_t}{\beta_t}, 
\label{bc_t}\label{wtd_comb_bf_vec_bs_ue}
\end{eqnarray}
where the normalization factors in \eqref{bc_r} and \eqref{bc_t} ensure that the beamforming vectors are of unit norm.
%$||\vec{w}_r^{(p)}||_2^2 = 1, \, \forall p$. Because of the scaling factor in \eqref{bc_t}, we have $||\vec{w}_{t_i}^{(q)}||_2^2 = \beta_t\beta_r$.
%Now, the total squared norm (power) of the transmit beamforming vectors used at every BS for the entire training phase is $PQ \beta_t\beta_r  = N_tN_r$, which is same as the total power of transmit beamforming vectors of the B-S method. This ensures some fairness in performance comparison, as the total power spent by both schemes for the training phase are identical, even though the training overheads are different. 

To obtain the received signal model for the beamforming vectors given in equation \eqref{wtd_comb_bf_vec_bs_ue}, we first note that,
\begin{align*}
\small{\vec{F}^*_{N_t}\vec{w}_{t}^{(q)}} & = \vec{F}_{N_t}^*\sum\limits_{l_t=1}^{\beta_t} \frac{1}{\sqrt{\beta_t}} [\vec{F}_{N_t}]_{:,(q-1)\beta_t + l_t} \\& = \frac{1}{\sqrt{\beta_t}} [\vec{I}_{\frac{N_t}{\beta_t}}(q) \otimes \vec{1}_{\beta_t}], %\\
%\small{\vec{w}_r^{(p)H}\vec{F}_{N_r}} & = \sum\limits_{l_r=1}^{\beta_r} \frac{([\vec{F}_{N_r}]_{:,(p-1)\beta_r + l_r})^H \vec{F}_{N_r}}{\sqrt{\beta_r}} \\&=  \frac{\small{[\vec{1}_{\beta_r} \otimes \vec{I}_{\frac{N_r}{\beta_r}}(p)]^T}}{\beta_r},
\end{align*}
where $\vec{1}_{k}$ represents an all one column vector of size $k \times 1$ and $\vec{I}_{k}(p)$ is the $p^{th}$ column of a $k \times k$ Identity matrix and $\otimes$ denotes the Kronecker product. We also have
\begin{align}
\vec{w}_r^{(p)*}\vec{F}_{N_r} = \frac{\small{[\vec{I}_{\frac{N_r}{\beta_r}}(p) \otimes \vec{1}_{\beta_r} ]^T}}{\sqrt{\beta_r}}. 
\end{align}
Based on these identities, the expression for the received symbols in \eqref{pq_meas_g} is given by,  %derived in the same procedure as in equation \eqref{simplified_obs_model_BS} as,
\begin{align}
y_{p,q} & =  \frac{\sqrt{\rho} x_{p,q}}{\sqrt{\beta_t \beta_r}} \sum\limits_{l_t=1}^{\beta_t}\sum\limits_{l_r=1}^{\beta_r}  [\vec{G}]_{(p-1) \beta_r + l_r, (q-1)\beta_t + l_t} + n_{p,q},
\label{simplified_obs_model_wtd_comb}
\end{align}
for $p =1,..., \frac{N_r}{\beta_r}$ and $q =1,..., \frac{N_t}{\beta_t}$. Intuitively, because of the widened beams used at both the BSs and UE, multiple DFT bins get mapped to the same scalar measurement at the UE. Specifically, the sum of all the elements of a $\beta_r \times \beta_t$ sub-matrix of $\vec{G}$ specified by the indices $\mathcal{S}_{p,q} = \{((p-1) \beta_r + 1, (q-1)\beta_t + 1), 
\cdots, (p \beta_r, q \beta_t) \}$ contributes to the single measurement $y_{p,q}$ at the UE. Now, the active set $\mathcal{S}_{\text{BC}}$ is defined as the set of indices $\{(p,q)\}$ for which the
corresponding observation $y_{p,q}$ in \eqref{simplified_obs_model_wtd_comb} has contribution from at least one non-zero channel entry from the support set $\mathcal{S}$ of the channel matrix $\vec{G}$, 
which is given in \eqref{support_set}. We have, $\mathcal{S}_{\text BC} = \{ (p,q) \Big | \mathcal{S}_{p,q} \cap \mathcal{S} \neq \emptyset \}$.

We employ the energy detector rule \eqref{thr_condition} to detect the presence of the active BS. If the index $(p,q)$ is detected, the DFT bin indices from the set $\mathcal{S}_{p,q}$ give 
the associated AoA-AoD pairs. For the BC technique, the resolutions of the detected transmit and receive beams are wider by a factor of $\beta_t$ and $\beta_r$ respectively, compared to the CBS scheme. 

\begin{theorem} \label{thm:bc}
Under ideal on-grid channel conditions \eqref{ideal1},\eqref{ideal2}, the false alarm probability for the BC scheme is
\begin{eqnarray}
\mathbb{P}_F &=& 1-\Big(1-e^{\frac{-\tau}{\sigma_n^2}}\Big)^{\frac{N_t N_r}{\beta_t \beta_r} - |\mathcal{S}_{\text{BC}}|},
\label{thres_expression_wtd_comb} 
\end{eqnarray}
The probability of detecting the BS is lower bounded as 
\begin{align}
\mathbb{P}_D &\geq \exp \Big(-\frac{\tau}{\sigma_n^2 + \frac{\rho}{\beta_t \beta_r} \max_k \sigma_{k}^2} \Big).
\label{Pd_BC}
\end{align}
\end{theorem}

\begin{proof}
Under ideal channel assumptions, out of the $\frac{N_t N_r}{\beta_t \beta_r}$ observations in \eqref{simplified_obs_model_wtd_comb}, only symbols associated with $\mathcal{S}_{\text{BC}}$ 
have contributions from the non-zero channel gain. Rest of the observations are purely i.i.d. Gaussian noise. The false alarm probability for the BC scheme (when at least one of the noise samples crosses the threshold) is obtained 
in the same manner as in \thmref{cbs}.  
%Once again, under \emph{idealized channel assumptions}, the measurements $y_{p,q},\, \forall p,q$ are independent but not identically distributed. Due to which, we cannot derive an exact expression for $\mathbb{P}_D$, but a lower bound can be found by following the same procedure adopted to prove \textbf{Theorem 1} - \textbf{Overlap case}.
%\begin{lemma}
%Suppose the training beamforming vectors are chosen according to equation \eqref{wtd_comb_bf_vec_bs_ue}, and the pilot symbols transmitted are such that $x_{p,q}^{(i)} = x, \forall p = 1, ..., \frac{N_r}{\beta_r}, \forall q = 1, ..., \frac{N_t}{\beta_t}, \forall i = 1, ..., N_{\text{bs}}$, then under the \textit{idealized channel assumption}, the probability of successful detection ($\mathbb{P}_D$) can be lower bounded as in equation \eqref{Pd_RBS} where the threshold $\tau$ is given by,
%where $\mathcal{S}_{\text{actv}} = \Big\{(p,q) \Big| y_{p,q} \, \text{is generated by} \, [\vec{G}_i]_{\bar{p},\bar{q}}, $ for some $ (\bar{p},\bar{q}) \in \mathcal{S} \Big\}$
%\end{lemma}
%\begin{proof}
Detection happens if any one of $y_{p,q}$ for $(p,q) \in \mathcal{S}_{\text{BC}}$ crosses the energy threshold $\tau$ in \eqref{thr_condition}. Let $\sigma^2_{p,q}$ 
denote the signal variance of $y_{p,q}$, which is defined as the variance of the $\frac{\sqrt{\rho} x_{p,q}}{\sqrt{\beta_t \beta_r}} \sum\limits_{l_t=1}^{\beta_t}\sum\limits_{l_r=1}^{\beta_r}  [\vec{G}]_{(p-1) \beta_r + l_r, (q-1)\beta_t + l_t}$. We note that $\max_{(p,q)} \sigma^2_{p,q} \geq \frac{\rho}{\beta_t \beta_r} \max_k \sigma_k^2$. At least one entry $y_{p,q}$ has signal variance      
$\sigma^2_{p,q} \geq \frac{\rho}{\beta_t \beta_r} \max_k \sigma_k^2$. Hence the detection probability is lower bounded by the probability of a Gaussian random variable with variance 
$\frac{\rho}{\beta_t \beta_r} \max_k \sigma_k^2$ crossing the energy threshold $\tau$, as given in \eqref{Pd_BC}.
\end{proof}
\subsection{Analysis on CD failure \& Time complexity}
\par Cell search in general is not a one-time attempt. BSs transmit pilot signals at regular time instants to provide chances for UEs to get connected to them. We declare CD failure only if the UE is not able to detect at least one active BS within $N_{max}$ attempts. $N_{max}$ depends on how fast the channel varies and is directly related to the coherence time of the channel. The multi-path arrival and departure angles vary slowly when compared with the channel gains. We assume  the AoA-AoD of the channels remain same (thus $\mathcal{S}$) and only the channel gain varies during $N_{max}$ attempts. We use probability of CD failure, $\mathbb{P}_{fail}$, to quantify the CD failure, and characterize it in \thmref{P_fail}.

\begin{theorem} \label{thm:P_fail}
Under ideal on-grid channel conditions, the probability of CD failure for the CBS scheme can be written as,
\begin{equation}
\mathbb{P}_{fail} = (1-\mathbb{P}_D)^{N_{max}}
\end{equation}
where expression for $\mathbb{P}_D$ is given in \eqref{Pd_BS}. Similarly, lower bound on $\mathbb{P}_{fail}$ can be derived for BC scheme using the upper bound on $\mathbb{P}_D$ (obtained in equation \eqref{Pd_BC}).  
\end{theorem}. 
\vspace{-8mm}
\begin{proof}
$\mathbb{P}_{fail}$ is the probability of CD failure in all the $N_{max}$ repeated trials. Since the statistics of the channel paths remain unchanged for every trial, each trial will have a probability of detecting at least an accessible BS equal to $\mathbb{P}_D$. Further, every trial is independent of another. Applying these results, we obtain the expression as in \thmref{P_fail}. It is clear from the expression that $\mathbb{P}_{fail}$ decreases with increase in $\mathbb{P}_D$ and $N_{max}$. As $N_{max} \rightarrow \infty$, it means that the trial is repeated a large number of times, and $\mathbb{P}_{fail} \rightarrow 0$ 
\end{proof}
\par A measure of time complexity that is closely related with $\mathbb{P}_{fail}$ is the average number of attempts ($N_{avg}$) made for the first detection, within the maximum number of attempts fixed as $N_{max}$. $N_{avg}$ for any scheme will be analogous to its $\mathbb{P}_D$ value and will decrease with increase in $\mathbb{P}_D$. $N_{avg}$ for CBS and BC schemes are characterized in \thmref{N_avg}.
\begin{theorem} \label{thm:N_avg}
With $\mathbb{P}_D$ defined as in \thmref{cbs} for CBS scheme, for a fixed $\mathbb{P}_F$, $N_{avg}$ within a maximum of $N_{max}$ attempts can be derived as,
\begin{equation}
    N_{avg} = \frac{1}{\mathbb{P}_D} - \frac{N_{\text{max}}(1-\mathbb{P}_D)^{N_{\text{max}}}}{1-(1-\mathbb{P}_D)^{N_{\text{max}}}}
\end{equation}
\end{theorem}
\begin{proof}
Let $N$ be the random variable indicating the number of trials required to see the "first detection"and let $Q$ be the event that $N \leq N_{max}$. Then, $N$ follows geometric distribution and the probability mass function (pmf) of $N = n$ is given by $\mathbb{P}(N=n)=(1-\mathbb{P}_D)^{n-1}\mathbb{P}_D, \forall n$ where $\mathbb{P}_D$ is the CD probability of detecting at least one BS in an attempt. Probability of the event $Q$ is $1-(1-\mathbb{P}_D)^{N_{max}}$. Conditional pmf of the number of attempts given the event $Q$ can be then derived as,
\begin{align*}
    \mathbb{P}(N =n \mid Q) &=\frac{ \mathbb{P}(Q \mid N=n)\mathbb{P}(N=n)}{\mathbb{P}(Q)} \\
    &= \frac{1.(1-\mathbb{P}_D)^{n-1}\mathbb{P}_D}{{\mathbb{P}(Q)} }, 1\leq n \leq N_{max}
\end{align*}
Now, avg. number of trials needed will be,
\begin{align*}
N_{{avg}} = \mathbb{E}[N \mid Q] &= \sum_{k=1}^{N_{max}} k \mathbb{P}(N =k \mid Q) \\ &= \sum_{k=1}^{N_{{max}}} k \frac{\mathbb{P}_D (1-\mathbb{P}_D)^{k-1}}{\mathbb{P}(Q)} \\& =  \frac{1}{\mathbb{P}(Q)}\Big({\mathbb{P}_D} - N_{max}(1-\mathbb{P}_D)^{N_{max}} \Big) \\& = \frac{1}{\mathbb{P}_D} - \frac{N_{max}(1-\mathbb{P}_D)^{N_{max}}}{1-(1-\mathbb{P}_D)^{N_{max}}}.
\end{align*}
\end{proof}
Substituting $\mathbb{P}_D$ expressions in \thmref{cbs} and \thmref{bc}, analytical expression and bound for $N_{avg}$ can be obtained for CBS and BC schemes respectively. Note that, as $N_{max}$ becomes large, the second term in the $N_{avg}$ expression vanishes, and $N_{avg} \approx \frac{1}{\mathbb{P}_D}$ which is the expectation of the geometric distribution with parameter $\mathbb{P}_D$. 

\section{Beam Sweep with Sequence transmission}
\label{sec:sequence transmission}
%\subsection{System Model}
\par In many communication systems (a cellular network, for example), there are multiple base stations (BSs) in the network, with each BS given a unique identity.
In the cell discovery problem with multiple BS, the UE needs to find the presence of at least one suitable BS from which the UE has sufficient received signal strength to establish a communication link. In addition, the UE also needs to find the identity of the detected BS and the corresponding AoA-AoD pairs. Each BS is assigned a unique sequence in a cellular network, referred to as a synchronization sequence (SS), based on its identity. Each BS periodically transmits its unique synchronization sequence, in order to facilitate cell discovery. 
In 5G-NR standards, the mm wave systems are envisioned to transmit synchronization sequence along with directional beamforming \cite{Giordani_ICST_2019}. 
In this Section, we consider multiple BS and present details of the beam sweeping along with the transmission of SS. We analyze the detection performance of the energy detector for orthogonal SS, under on-grid
channel assumptions. 

\subsection{Synchronization Sequences}
\label{sec:sync sequences}
Consider a network with $N_B$ unique base stations, with their identities from the set $\mathcal{I} = \{1,\cdots,N_B\}$. Let $\mathcal{X}= \{\vec{x}_1,\cdots,\vec{x}_{N_B}\}$, each $\vec{x}_m$ being an 
$N_Z \times 1$ vector, denote the set of synchronization sequences (SSs) used by the BSs (BS with identity $m \in \mathcal{I}$ is assigned a SS $\vec{x}_m$). Orthogonal sequences are good candidates for synchronization sequences. In this case, $\vec{x}_m^* \vec{x}_n = 0$ for any $m,n \in \mathcal{I}$ with $m \neq n$. However, the orthogonal SS require that $N_Z \geq N_B$. In a network with large number of BSs (say, a cellular network), the requirement $N_Z \geq N_B$ is undesirable due to the 
large overhead for transmitting SS. Hence, non-orthogonal signals with small cross correlation are used as SSs. We discuss some of the SSs used in existing wireless standards.

Zadoff-Chu (ZC) sequences are commonly used as synchronization signals due to their good correlation properties  \cite{Hyder_2017_TWC}.
Let $\vec{s}_0, \vec{s}_1, ..., \vec{s}_{L-1}$ be root ZC sequences (with norm as $\sqrt{L}$) of length $L$ with $L$ being an odd prime number.
The $k^{\text{th}}$ value of $r^{\text{th}}$ root ZC sequence is given as,
\begin{align*}
    s_r(k)=e^{-i\pi r \frac{k(k+1)}{L}},0 \leq k \leq L-1
\end{align*}

 If $\vec{s}_{r}^{(l)}$ denote the $l^{th}$ cyclic shift of $\vec{s}_{r}$ with $l = 0,1,...,L-1$ and $r = 1,...,L-1$, then
\begin{enumerate}
\item $\Big(\textbf{s}_{r}^{(l_1)}\Big)^*\Big(\textbf{s}_{r}^{(l_2)}\Big) = 0$, for all $l_1 \neq l_2$.
\item $\Big(\textbf{s}_{r_1}^{(l)}\Big)^*\Big(\textbf{s}_{r_2}^{(l)}\Big) = \sqrt{L}$, for all $r_1 \neq r_2$ and $r_1, r_2 = 1,...,L-1$.
\end{enumerate}
The cyclic shifts of the root sequences can also be considered as ZC sequences. The first sequence $\vec{s}_0$ is an all-one sequence $\vec{1}_L$, and the shifted versions cannot be used as SS.  Hence, in total, we can have $L(L-1)+1$ ZC sequences of odd prime length $L$ with good correlation properties.
Gold codes can also be used for synchronization purposes because of their good cross-correlation properties \cite{Aditi_TSP_2020}. 
%The upper limit $r(n)$ on the cross-correlation value of shifted Gold sequences of length $L=2^n-1$ is,
%\begin{align*}
%    r(n) & = \frac{2^{\frac{(n+1)}{2}}+1}{L},\text{if $n$ is odd} \\
 %   & = \frac{2^{\frac{(n+2)}{2}}+1}{L},\text{if $n$ is even}
%\end{align*}
In this paper, we use ZC sequences as synchronization signals for multiple BS case. 

\subsection{Beam Sweep with Synchronization Signals}

\par We describe the conventional beam sweeping technique coupled with SS transmission. In this scheme, each BS transmits its assigned SS in all possible AoA-AoD beam direction pairs. Let $\mathcal{A} \subset \mathcal{I}$ denote the set of \emph{active} BSs from which the given UE receives the signal. Let  $N_A = |\mathcal{A}|$ denote 
the number of active BS and $N_B$ denote the total number of BSs. We assume that each BS has a ULA of size $N_t$. Let  $\vec{H}_i$ denote the channel matrix between the 
UE and the BS with identity $i$. If $i$ corresponds to an active BS, that is, $i \in \mathcal{A}$, then $\vec{H}_i$ follows the model given in \secref{chan}. If $i$ is not an active BS, then $\vec{H}_i = \vec{0}$. 
The received signal model with multiple BSs is,
\begin{align}
y &= \vec{w}_r^* \sum\limits_{i \in \mathcal{A}} \vec{H}_i \vec{w}_{t_i} \sqrt{\rho_i} x_i +n
\end{align}
where $\vec{w}_{t_i}$ is the beamforming weights of the BS with identity $i$, $x_i$ is the symbol transmitted by that BS and $\rho_i$ governs the received power level of the BS. The SNR for the
above reception model is given as,
\begin{align}
\text{SNR} &= \frac{\big|\vec{w}_r^* \sum\limits_{i \in \mathcal{A}} \vec{H}_i \vec{w}_{t_i} \sqrt{\rho_i} x_i \big|^2 }{\sigma_n^2}. 
\label{SNR_multipleBS}
\end{align}
In the training phase, each BS transmits its SS in each beam direction (AoA-AoD pair), by setting the transmit beamforming weights and receive beamforming weights as in CBS, with
\begin{align}
& \vec{w}_{t_i}^{(q)} = [\vec{F}_{N_t}]_{:,q}, \,\, q \in \{1, ..., N_t \}, \, i \in \mathcal{I}  
\label{1simplified_weight_vectors_1} %\\&
%\vec{w}_{r}^{(p)} = [\vec{F}_{N_r}]_{:,p}, \,\, p \in \{1, ..., N_r \}.
%\label{1simplified_weight_vectors_2}
\end{align}
and the receive beamforming weights as in \eqref{simplified_weight_vectors_2}. With $p \in \{1,\cdots,N_r\}, q \in \{1,\cdots,N_t\}, \ell \in \{1,\cdots,N_Z\}$, the training phase observations
are given by,
\begin{align}
y_{p,q,\ell} & =  [\vec{F}_{N_r}]_{:,p}^*  \sum\limits_{i\in\mathcal{A}} {\vec{H}_i} [\vec{F}_{N_t}]_{:,q}  \sqrt{\rho_i} x_{i,\ell} + n_{p,q,\ell} \nonumber \\ 
& =  \sum\limits_{i \in \mathcal{A}} \sqrt{\rho_i} [\vec{G}_i]_{p,q} x_{i,\ell} + n_{p,q,\ell},
\label{1simplified_obs_model_BS}
\end{align}
where $x_{i,\ell}$ denotes the $\ell^{th}$ sample of the SS $\vec{x}_i$. We assume that $|x_{i,\ell}| =1, \forall i, \ell$, which holds true for SS used in practice, including ZC sequences and Gold codes. We have a total of $N_t N_r N_Z$ of observations for the training phase. We correlate this set of observations with SS from each BS, in each beam direction (AoA-AoD pair) as 
\begin{align}
z_{p,q,m} = \sum\limits_{\ell = 1}^{N_Z} y_{p,q,\ell} x^*_{m,\ell}, \,\, m\in \{1,\cdots,N_B\}. \label{eq:cormet}
\end{align}
Note that $z_{p,q,m}$ denotes the correlation metric corresponding to the BS with identity $m$, in the beam direction given by the 2D DFT bin pair $(p,q)$. Corresponding energy detector rule is  
\begin{align} 
|z_{p,q,m}|^2 > \tau, \label{energy1}
\end{align}
where $\tau$ is a suitably chosen threshold.
In a similar manner, we can incorporate the SS transmission and detection with the beam combining scheme as well.
\subsection{Analysis of CBS with orthogonal SS}
In order to analytically characterize the $\mathbb{P}_F$ and $\mathbb{P}_D$ of the energy detector in \eqref{energy1}, we make some additional assumptions. As before, we 
consider on-grid channels for active BS such that, for $i \in \mathcal{A}$, $\vec{G}_i$ has exactly $K_i$ independent Gaussian entries with variances $\sigma_{k,i}^2$. We also 
assume that SS of different BS are mutually orthogonal, such that $\vec{x}_i^* \vec{x}_m = 0$, for $i \neq m$.  
Under these assumptions, we have 
\begin{align}
z_{p,q,m} &= \sum\limits_{\ell = 1}^{N_Z} \left(\sum\limits_{i \in \mathcal{A}} \sqrt{\rho_i} [\vec{G}_i]_{p,q} x_{i,\ell} + n_{p,q,\ell} \right) x^*_{m,\ell} \\
&= \sum\limits_{i \in \mathcal{A}} \sqrt{\rho_i} [\vec{G}_i]_{p,q} \sum\limits_{\ell = 1}^{N_Z} x_{i,\ell} x^*_{m,\ell}  + 
\underbrace{\sum\limits_{\ell = 1}^{N_Z} n_{p,q,\ell} x^*_{m,\ell}}_{\tilde{n}_{p,q,m}}. \label{cormet2}
\end{align}
Due to the orthogonality and unit modulus assumptions on SS, we have 
\begin{align}
\sum\limits_{\ell = 1}^{N_Z} x_{i,\ell} x^*_{m,\ell} &= \begin{cases} 0 & \,\,  i\neq m, \\ N_Z & \,\, i = m, \end{cases}
\end{align}
and hence $z_{p,q,m} = \sqrt{\rho_m} N_Z [\vec{G}_m]_{p,q} + \tilde{n}_{p,q,m}.$
Note $\tilde{n}_{p,q,m}$ are i.i.d. Gaussian with variance $N_Z \sigma_n^2$. Also, note that $[\vec{G_m}]_{p,q}$ is non-zero only if $m \in \mathcal{A}$ and 2D DFT bin pair $(p,q)$ corresponds
to one of the non-zero multipaths gains in $\vec{G}_m$. Out of the $N_t N_r N_Z$ correlation metrics $\{z_{p,q,m}\}$, only $K_{\text{tot}} = \sum_{i \in \mathcal{A}} K_i $ of the metrics have non-zero channel gains. Rest of the metrics are purely i.i.d. Gaussian noise. Based on this observation, we have the following theorem, which can be proved in the same manner as \thmref{cbs}.
\begin{theorem} \label{thm:cbsss}
For the conventional beam sweep technique with orthogonal synchronization transmission, under the on-grid channel conditions,
the probability of false alarm is given by
\begin{equation}
\mathbb{P}_F = 1-\Big(1-e^{\frac{-\tau}{N_Z \sigma_n^2}}\Big)^{N_t N_rN_B - K_{\text{tot}}},
\end{equation}
and the probability of successful detection $\mathbb{P}_D$ is given by 
\begin{align}
\mathbb{P}_D = 1-  \prod \limits_{i \in \mathcal{A}} \prod \limits_{k=1}^{K_i} \Big[1-\exp \Big(-\frac{\tau}{N_Z \sigma_n^2 + \rho_i N_Z^2 \sigma_{k,i}^2} \Big)\Big].
\label{Pd_BS_multipleBS}
\end{align}
\end{theorem}

\section{Simulation Results}
\label{sec:Num_result}
In this section, we present our simulation results evaluating the performance of CBS and BC cell discovery algorithms in terms of probability of successful detection $\mathbb{P}_D$, probability of CD failure $\mathbb{P}_{fail}$, and average number of attempts for CD success $N_{avg}$ with a constraint on the probability of false alarm $\mathbb{P}_F$.

\subsection{Simulation Setup \& Channel Generation:}

%\begin{figure}
%\center
%\includegraphics[width = 8cm, height = 6.5cm]{PdvsSNR_Nt32Nr8Nbs1Nbstot1_idealchannel_CBS.eps}
%\caption{Comparison of analytical and simulated $\mathbb{P}_D$ along with NYUSIM results for CBS.}
%\label{fig:CBS_PD_expression_verify}
%\vspace*{-5mm}
%\end{figure}
\begin{figure*}[t]
\center
\includegraphics[width = 16.5cm, height = 6cm]{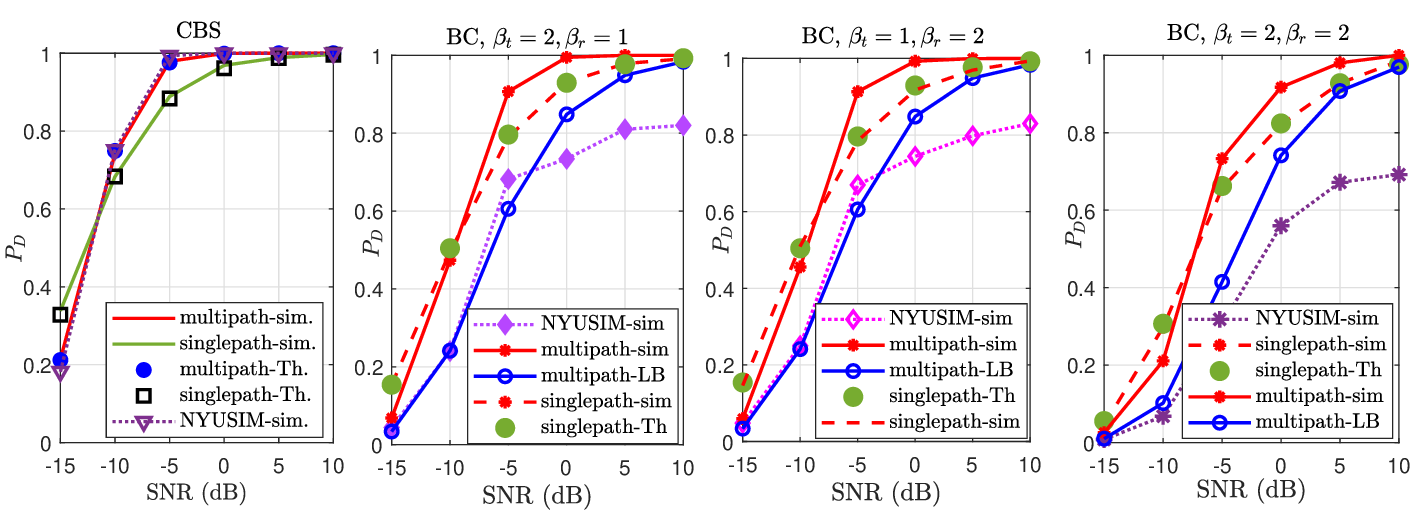}
\caption{Comparison of the analytical  and simulated $\mathbb{P}_D$ along with NYUSIM results for CBS and BC.}
\label{fig:BC_PD_expression_verify}
\vspace*{-5mm}
\end{figure*}

For simulation, we consider the $N_B$ number of BSs to be randomly deployed around the UE with in a radial distance of $R$ meters. Out of these BSs, we randomly select a set of $N_A$ active  BSs from which  the signals reach the given  UE.  The remaining inactive  BSs are assumed to have zero channel gain with the UE.
\par We generate the mm wave channel matrices using NYUSIM simulator (Version 1.6) \cite{Rappaport_2017_ICC,Rappaport_2017_VTC} which is an experimentally driven spatial channel simulator for mm wave communication systems. The simulator has been developed based on extensive real-world channel measurements and can be used for carrier frequencies ranging from $500$ MHz to $100$ GHz, and RF bandwidths from $0$ to $800$ MHz. The channels between the active BSs and the UE obtained, in this case, are not \emph{on-grid channels}. Further, the channels can be LoS or NLoS, and are determined based on the NYUSIM LoS probability model discussed in \cite{Rappaport_2017_VTC}.
We set the operating carrier frequency as $28$ GHz, set-up as "Urban Micro scenario" with $R=50$ and system bandwidth $= 5$ MHz. All other parameters in the NYUSIM GUI are set to their default values. Note that, even though the system bandwidth may be large, the training signals for cell discovery can be transmitted over a smaller band. For instance, in LTE standards, the synchronization signals are transmitted over a bandwidth of $1$ MHz, while the system bandwidth can be up to $20$ MHz. Further, the Frobenius norm of channel matrix between any BS and UE is normalized to 1. The transmit power levels $\rho_i$s used at different BSs also result in  variation in their respective received signal strength at the UE. We fix $\rho_i \forall i$ as a random number between 1 to 10. Hence, the maximum difference between the power levels of the strongest and weakest active BS will be 10dB.

We set the non-zero channel entry threshold in \eqref{support_set} as $\delta = \frac{1}{2}\Big(\max\limits_{a,b} \Big|[ \vec{G}_i]_{a,b} \Big|^2\Big)$, that is, any entry within $3$ dB of the absolute square of the largest magnitude entry in $\vec{G}_i$ is considered a non-zero entry. The threshold $\tau$ in the detector \eqref{thr_condition} and \eqref{energy1} is set as $\kappa \sigma_n^2$, where the $\kappa$ is chosen such that 
the probability of false alarm $\mathbb{P}_F$ (which is the probability of identifying an inactive BS and/or identifying incorrect AoD-AoA pair) is at most $0.01$. Note that the value of $\kappa$ in general will be different for different techniques.

\subsection{Verifying the derived $\mathbb{P}_D$ expressions using simulations:}

\par We verify the $\mathbb{P}_D$ expressions derived for the CBS and BC schemes using simulations in \Figref{BC_PD_expression_verify}. We plot the $\mathbb{P}_D$ curves for \emph{ideal channel} set-up, as a function of SNR which is 
defined as, $\text{SNR} = \frac{\sum_{i =1}^{N_A}\sum_{k_i=1}^{K_i} \rho_i \sigma_{k_i}^2}{N_t  N_r \sigma_n^2}$ with channel gains distributed as $\mathcal{CN}(0,\sigma_{k_i}^2)$. Here, $\sigma_{k_i}^2 = \frac{N_tN_r}{K_i} \alpha_{k,_i} $ and $\alpha_{k_i} \in (0,1), \, \forall k_i$. We fix $N_t = 32, N_r = 8$ and $N_A = N_B = 1$. We consider multi-path scenario with $K=3$ and for single-path scenario, $K$ is set to $1$. 
%Using this expression, we calculate the noise variance to meet a particular received SNR at the UE for any channel realisation, and generate the AWGN accordingly.
%\underline{(Do we need the previous sentence saying how we generate noise vector?).
From \Figref{BC_PD_expression_verify}(a), we observe that for \emph{ideal channel} conditions both analytical and simulation results exactly match for the CBS scheme. In addition, we also plot the $\mathbb{P}_D$ performance obtained using the NYUSIM channels (which are not on-grid) with the SNR defined in \eqref{snr1}, and observe the close similarity with the performance of multipath \emph{ideal channel} set-up. We analyze the analytical and simulation results for BC scheme with different $\beta_t$ and $\beta_r$ values in \Figref{BC_PD_expression_verify}(b), \Figref{BC_PD_expression_verify}(c) and \Figref{BC_PD_expression_verify}(d). BC scheme uses widened/combined beams for exhaustive search and hence has reduced training overhead ($M=128$; $M=64$) compared to the CBS scheme (which requires $M=256$). From \eqref{Pd_BC}, the lower bound for $\mathbb{P}_D$ is met with equality when the channel contains a single path (i.e., $K = 1$). This is also verified through simulation in \Figref{BC_PD_expression_verify}. 
\par
\begin{figure}
\center
\includegraphics[width = 7.75cm, height = 6.5cm]{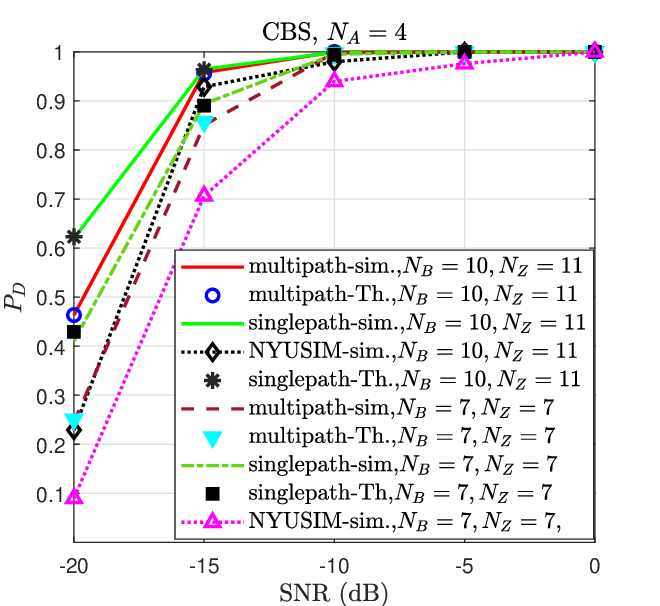}
\caption{Comparison of the analytical $\mathbb{P}_D$ expressions with NYUSIM results for BS with sequence transmission.}
\label{fig:seqtxn_PD_expression_verify}
\vspace*{-2mm}
\end{figure}

 We plot the $\mathbb{P}_D$ performance of CBS  with orthogonal sequence  based  transmissions in \Figref{seqtxn_PD_expression_verify} with respect to SNR as defined in \eqref{SNR_multipleBS}. We set $N_t=  32,N_r=  8, N_A= 4  \text{ and } K_i - \{3,4,2,2\}$ for $i \in A$. Here too, we see that the analytical \eqref{Pd_BS_multipleBS} and simulation results match for the \emph{ideal channel condition}. We verify the same for $N_B=10,N_Z=11$ case and $N_B=7,N_Z=7$ case. Also, analytical $\mathbb{P}_D$ plot closely matches the performance obtained using mm wave channels generated with NYUSIM.

\subsection{Performance comparison of CBS and BC}
\par
\begin{figure}
\center
\includegraphics[width = 7.75cm, height = 7.5cm]{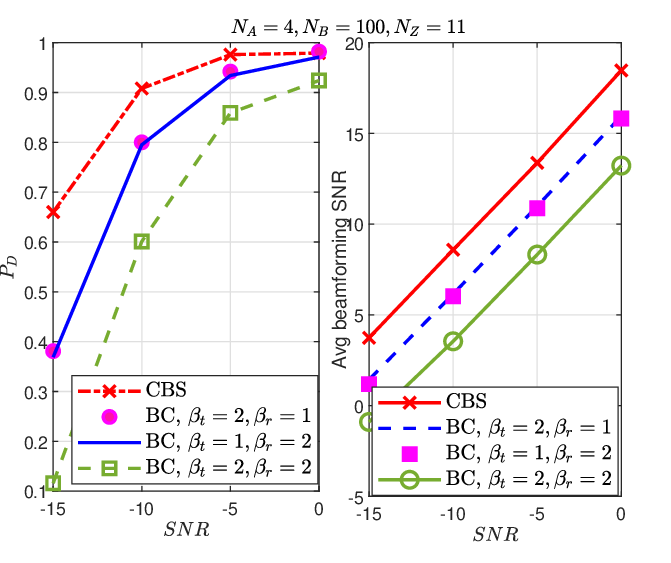}
\caption{(a) Detect at least 1 active BS using sequence transmission. (b)Average beamforming SNR for the detected BS}
\label{fig:seq txn plot}
\vspace*{-5mm}
\end{figure}

 We plot $\mathbb{P}_D$ vs SNR (defined in \eqref{SNR_multipleBS}) for mm wave channels generated using NYUSIM in \Figref{seq txn plot}(a). We fix $N_t=32,N_r=8, N_B=100, N_A=4 \text{ and } N_Z=11$. Being an exhaustive search mechanism, the CBS scheme out-performs BC but at the expense of large training overhead ($M= 256$). BC scheme uses widened/combined beams for exhaustive search and hence has reduced training overhead. As we increase the overall reduction factor $\beta_r \beta_t$, the resolution of the estimated AoA-AoD decreases, affecting the detection performance for the BC method.
\par After CD, let the detected BS identity be $i_0$ and the corresponding AoA-AoD pair be $(p_0,q_0)$.
Now, we plot the \emph{average beamforming SNR} achieved using the beamforming vectors 
directed to the detected BS $i_0$ with the detected AoA-AoD pair $(p_0,q_0)$ for various training schemes. Specifically, with $\vec{w}_t = [\vec{F}_{N_t}]_{:,q_0}$ and $\vec{w}_r = [\vec{F}_{N_r}]_{:,p_0}$ being the beamforming directions, we define the beamforming SNR as, $\frac{\Big| \sqrt{\rho_{i_0}}~\vec{w}_r^H \vec{H}_{i_0} \vec{w}_t\Big|^2}{\sigma_n^2}$. 
We average this beamforming SNR over multiple channel realizations under the condition that the detected BS is an active one. 

\par \Figref{seq txn plot}(b) shows the variation of average beamforming SNR with respect to received SNR for different CD schemes, given successful detection. We notice that when SNR increases, the average beamforming SNR also increases for all methods. Further, the average beamforming SNR performance is the highest for CBS schemes. However, with an increase in $\beta_r \text{ or } \beta_t$, the resolution of the estimated AoA-AoD pair decreases, resulting in a decrease of beamforming gain and beamforming SNR. 

\subsection{Effect of $N_B$}
\par
\begin{figure}
\center
\includegraphics[width = 8cm, height =7.5cm]{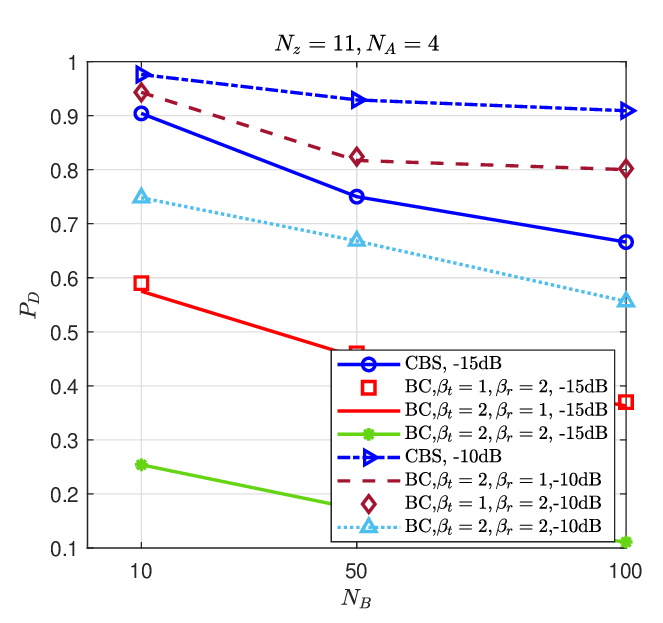}
\caption{Impact of total number of BSs in the network on $\mathbb{P}_D$ with NYUSIM results}
\label{fig:PDvsno_ofBSs}
\vspace*{-5mm}
\end{figure}

 In this section, we try to analyze CD performance in terms of the total number of BSs in the network ($N_{B}$). We plot $\mathbb{P}_D$ vs $N_B$ in \Figref{PDvsno_ofBSs}. We observe that $\mathbb{P}_D$ decreases as $N_{B}$ increases for all the schemes. As $N_B$ increases, UE has to correlate with the SS of all these BSs for CD. Thus, the threshold to maintain the fixed $\mathbb{P}_F$ ($0.01$) also increases resulting in the decrease of detection probability.

\subsection{Effect of $N_Z$}
\par
\begin{figure}
\center
\includegraphics[width = 8cm, height = 7cm]{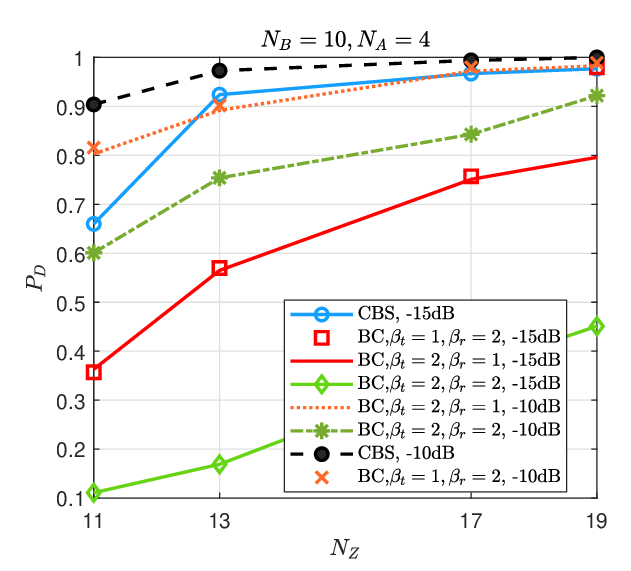}
\caption{Impact of the length of SS ($N_Z$) on $\mathbb{P}_D$ with NYUSIM results}
\label{fig:PDvsNz}
\vspace*{-5mm}
\end{figure}

 In \Figref{PDvsNz}, we study the variation of $\mathbb{P}_D$ with the length of synchronization sequence, $N_Z$. Here, we fix $N_B = 100$. The other parameters are: $N_t=32,N_r=8$ and $N_A=4$. We notice that for all the schemes, $\mathbb{P}_D$ increases with $N_Z$. This is because as $N_Z$ increases, the cross correlation between the sequences decreases. And as the correlation decreases, the sequences tend to become almost orthogonal which therefore increases the detection probability. 

\subsection{Analysis on CD failure \& Time complexity}
\begin{figure}
\center
\includegraphics[width = 7.75cm, height = 6.5cm]{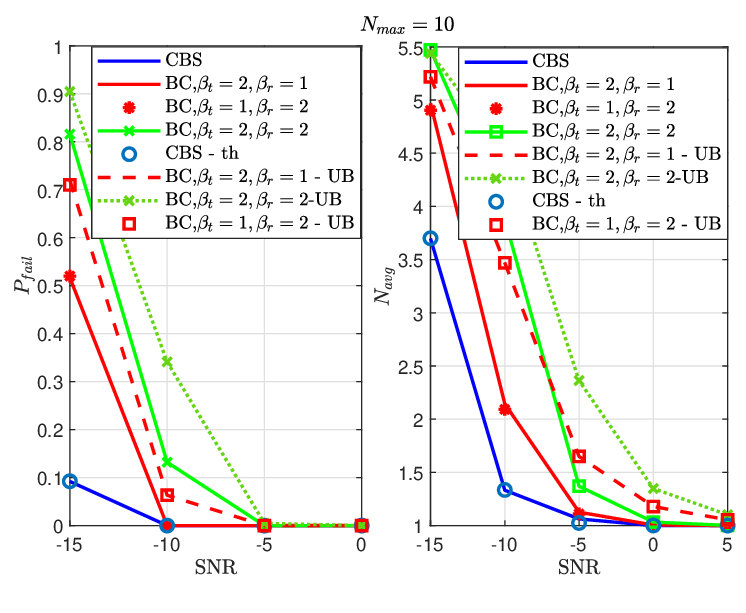}
\caption{(a) CD failure probability. (b) Time complexity in terms of  expected number of average attempts.}
\label{fig:P_fail and N_avg plot}
\end{figure}

\par In \Figref{P_fail and N_avg plot}, we quantify the efficiency of CBS and BC methods, in terms of ($\mathbb{P}_{{fail}}$) and the expected number of attempts ($N_{{avg}}$) it takes for a typical UE to establish a reliable connection, within a total of $N_{max}$ attempts. We assume \emph{ideal channel} set-up with $N_A=N_B=1,K=3 \text{ and } N_{max}=10$. The results are averaged over $1000$ channel realizations.  \Figref{P_fail and N_avg plot}(a) plots $\mathbb{P}_{{fail}}$ vs receive SNR. We observe that $\mathbb{P}_{{fail}}$ decreases with increase in SNR. CBS technique has the least $\mathbb{P}_{{fail}}$ because of its high training overhead and high beam resolution. \Figref{P_fail and N_avg plot}(b) plots $N_{{avg}}$ given successful detection of at least one active BS w.r.t SNR, thereby comparing the time complexity of the CD techniques. For both CBS and BC schemes, $N_{{avg}}$ decreases with increase in SNR. It is imperative to note that, higher the $\mathbb{P}_{{fail}}$ value for a CD scheme, the UE needs more attempts to establish a link with at least one active BS using that CD scheme, i.e., $N_{{avg}}$ will be larger. 

\section{Conclusion}
\par In this article, we addressed the beam sweeping techniques for the cell discovery problem in millimeter wave communication systems. We considered  beam sweep and beam combining schemes and presented the complete details of the transmit/receiver beamforming vectors in 
the training phase. We obtained analytical expressions of the detection probability, average number of CD trials, and probability of CD failure of the schemes under ideal channel assumptions. We also discussed the beam sweeping scheme with synchronization sequences when multiple BSs are present in the network and derived the expression for probability of detection when BSs transmit orthogonal synchronization sequences. We also presented simulation results using channels generated from NYUSIM for all the schemes.

\bibliographystyle{ieeetr}
\bibliography{macros_abbrev,beamforming_ref}

\vspace{10mm}

\begin{wrapfigure}{l}{25mm} 
\vspace{-5mm}
    \includegraphics[width=1in,height=1.25in,clip,keepaspectratio]{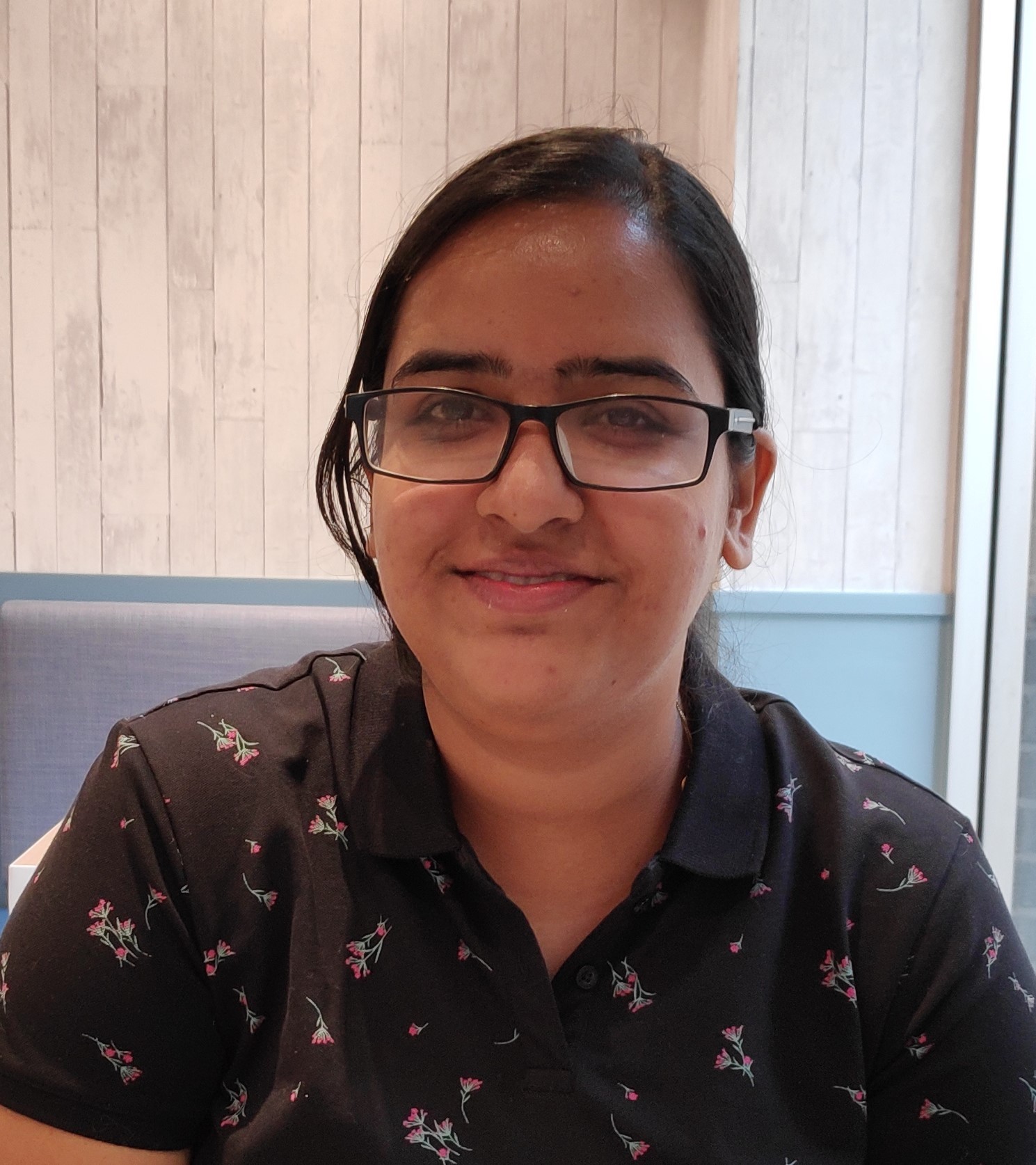}
  \end{wrapfigure}
 \textbf{Rashmi P} received the B.Tech. degree from MES College of Engineering, University of calicut, India, in 2011. She completed M.Tech in Signal Processing from National Institute of Technology, Calicut  in 2014, and is currently pursuing Ph.D. with Dept. of Electrical Engineering, IIT Madras, India.
 
\vspace{10mm}

\begin{wrapfigure}{l}{25mm} 
\vspace{-5mm}
    \includegraphics[width=1in,height=1.25in,clip,keepaspectratio]{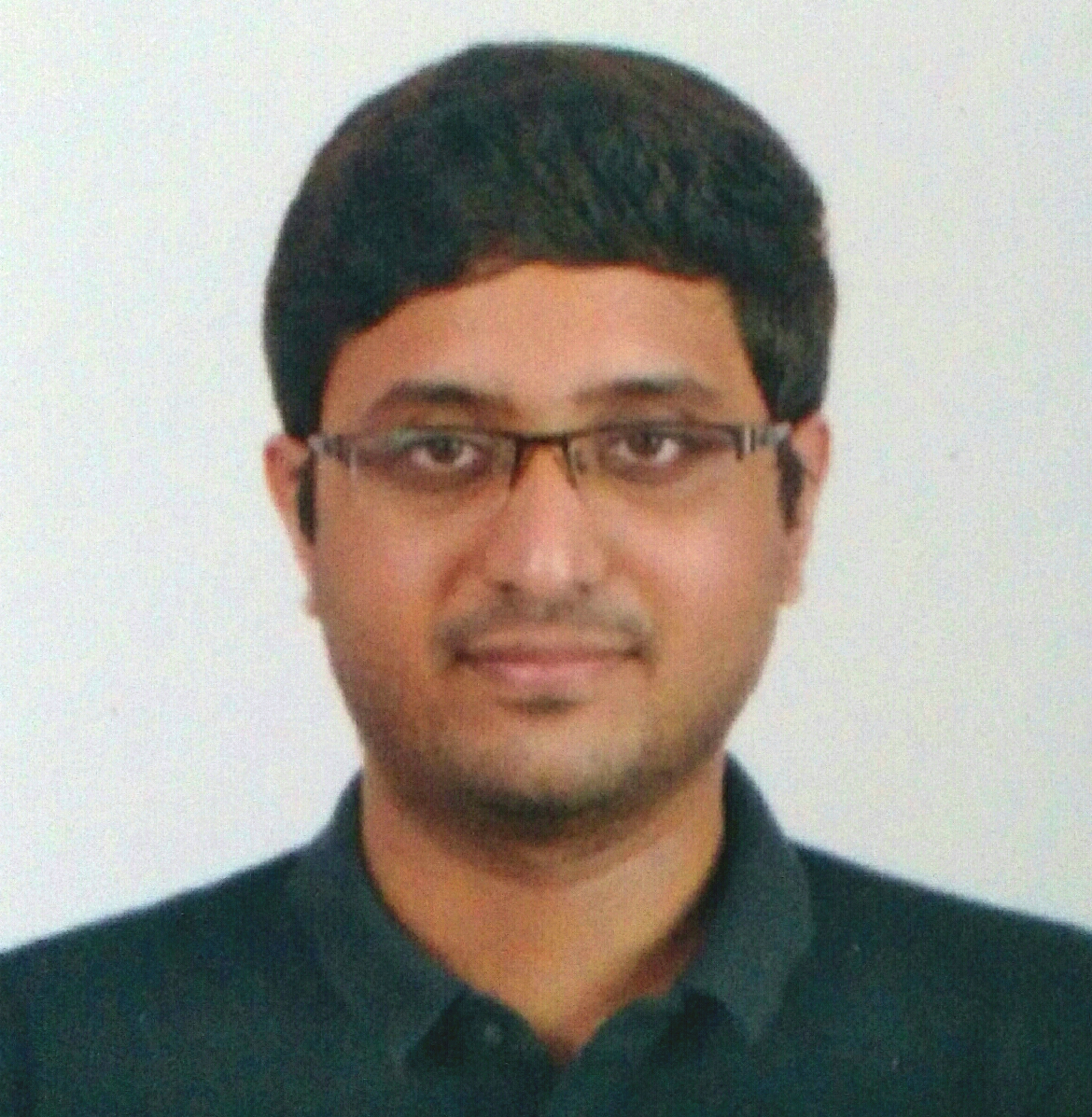}
  \end{wrapfigure}
 \textbf{ Manoj A} received the B.E. degree from the Rajalakshmi Engineering College, Anna University, Chennai, India, in 2014. He also completed M.S. and Ph.D. degrees with the Department of Electrical Engineering, IIT Madras, India in 2020.
 \\
 \vspace{8mm}
 
 \begin{wrapfigure}{l}{25mm} 
    \includegraphics[width=1in,height=1.25in,clip,keepaspectratio]{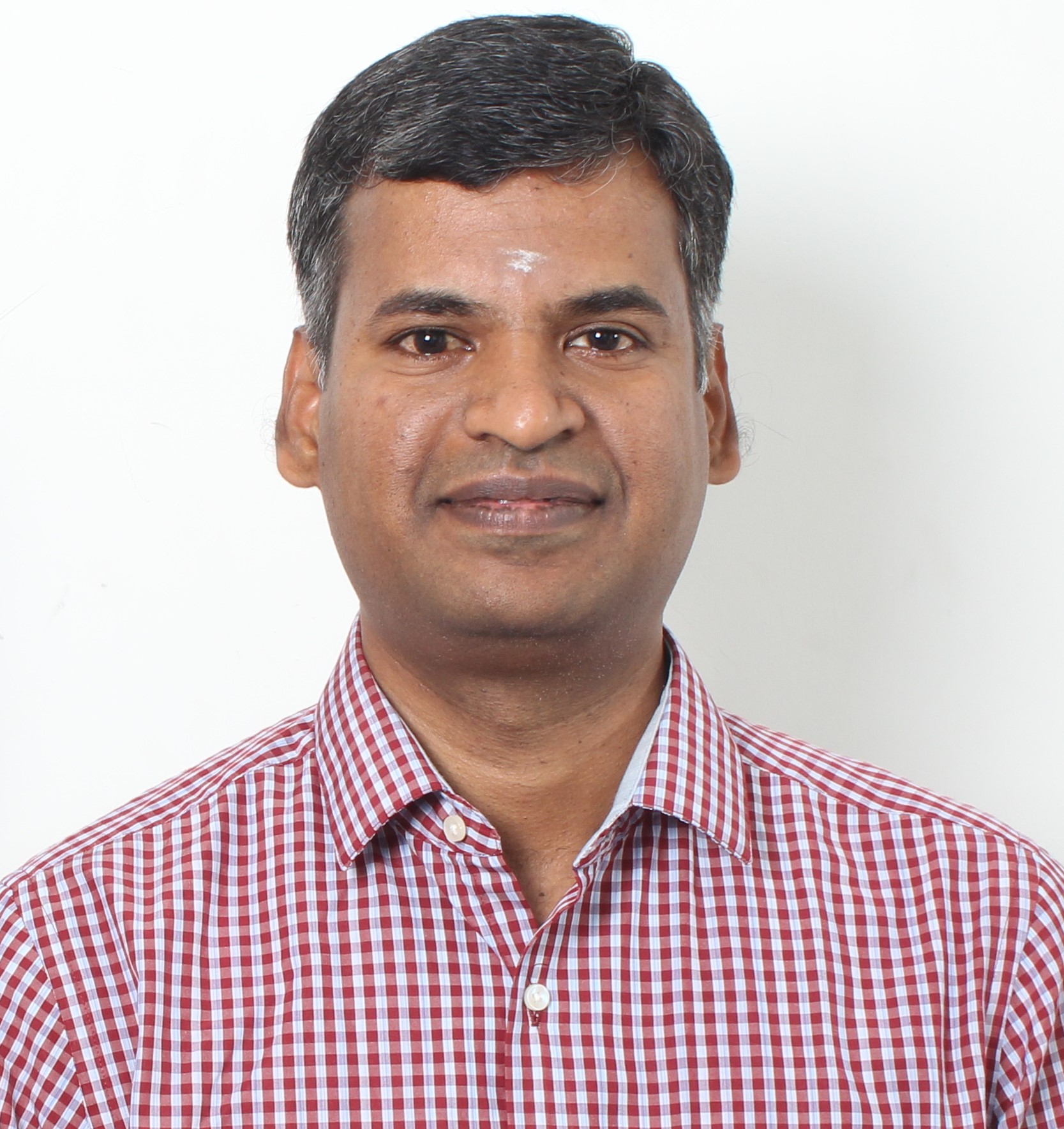}
  \end{wrapfigure}
 
\textbf{ Arun Pachai Kannu} received the B.E. (ECE) degree from the College of Engineering at Guindy in 2001 and the M.S. and Ph.D. degrees in electrical engineering from The Ohio State University in 2004 and 2007, respectively. From 2007 to 2009, he was a Senior Engineer with the Qualcomm Research Center, San Diego, CA, USA. He is currently an Associate Professor with the Department of Electrical Engineering, IIT Madras. His research interests include theory of sparse signal recovery and its applications in wireless communications.

\end{document}